\documentclass[12pt,preprint]{aastex}

\received{2001 November 6}
\begin{document}

\title{Long Term Spectral Variability of Seyfert Galaxies 
from RXTE Color-Flux Diagrams}

\author{I. E. Papadakis\altaffilmark{1,2},P. O. Petrucci\altaffilmark{3,4},
L. Maraschi \altaffilmark{3}, I. M. McHardy\altaffilmark{5}, P.
Uttley\altaffilmark{5}, and F. Haardt\altaffilmark{6}}

\altaffiltext{1}{IESL, Foundation for Research and Technology-Hellas, 
Heraklion, Crete, Greece; e-mail: jhep@physics.uoc.gr}

\altaffiltext{2}{Physics Department, University of Crete, 710 03,
Heraklion, Crete, Greece}

\altaffiltext{3}{Osservatorio Astronomico di Brera, Milano, Italy}

\altaffiltext{4}{Laboratoire d' Astrophysique de Grenoble, 38041,
Grenoble Cedex 9, France}

\altaffiltext{6}{Department of Physics and Astronomy, University of
Southampton, Southampton SO17 1BJ, UK}

\altaffiltext{6}{Universita dell'Insubria, Como, Italy}

\begin{abstract} 

We present results from $RXTE$ data obtained during a systematic
monitoring program of four Seyfert galaxies (NGC~5548, NGC~5506,
MCG~-6-30-15 and NGC~4051).  We studied the variability of three hardness
ratios derived from the light curves in four energy bands ($HR1$, which
describes the continuum variations, and $HR2$, $HR3$ which are sensitive
to the iron line and reflection component variations with respect to the
continuum, respectively). All the objects show similar spectral variations
in all ratios. In order to interpret the results we computed the hardness
ratios corresponding to a simple spectral model of a power law plus iron
line plus reflection component. In order to derive the model $HR2$ and
$HR3$ colors, we considered two possibilities: a) variations with constant
line equivalent width and reflection parameter $R$ (the case of a
reflecting/reprocessing material that responds with a short delay to the
continuum variability) and b) variations with constant line and reflection
flux (the case of a reprocessor that does not respond to the fast,
intrinsic variations).  The overall, mean observed trends can be explained
by spectral slope variations ($\Delta \Gamma \simeq 0.2-0.3$, and $\simeq
1$ for NGC 4051), and a constant flux Fe line and reflection component,
although the existence of a line component which is variable on short time
scales cannot be excluded. Finally, we find that the data are not
consistent with an increase of R with flux for individual sources,
indicating that, as a single source varies, softer spectra do not
correspond to larger $R$ values.

\end{abstract}


\section{Introduction}

X-ray emission is a ubiquitous property of Active Galactic Nuclei (AGN).
The spectral characteristics of this emission have been studied in detail
the past decade and our understanding of the X-ray emission properties has
improved considerably. In general, the X-ray spectra of AGN have a
power-law form over a large range in energy. The photon spectral index has
a typical value of $\Gamma \sim 2$ (Nandra \& Pounds 1994). Few sources
also show a high energy cutoff which can be modeled as an exponential
cutoff with e-folding energy of a few 100 keV (Gondek et al. 1996).
Furthermore, the presence of the 6.4 keV Fe line and the so-called
``Compton reflection hump" indicates the existence of cold, optically
thick material that reprocesses and reflects the medium energy X-rays. It
is generally believed that this material is located near the primary X-ray
source (i.e. the accretion disc). However, this is not the only
possibility. The cold material could also be in the form of a torus which
is surrounding and obscuring (depending on the inclination) the central
nucleus. The existence of this torus is predicted by models that unify
Seyfert 1 and 2 galaxies (e.g. Antonucci 1993). The contribution of the
torus to the spectrum of Seyfert 1 galaxies (by scattering part of the
primary radiation into the line of sight) can in principle explain the
reflection bump observed in the spectra of AGN (Ghisellini, Haardt \& Matt
1994).

Based on the detailed study of the X-ray spectra of the Seyfert 1 nuclei
the currently-popular scenario (the ``standard picture" say) is that
X-rays from AGN are produced by Compton upscattering of soft photons in a
hot ``corona" located above the accretion disc (e.g. Haardt \& Maraschi
1991, 1993). The X-ray continuum illuminates the disk, which is assumed to
be neutral, producing optical/UV emission via reprocessing, the iron line
and the reflection hump. Apart from the energy spectra, the X-ray flux and
spectral temporal variations can also constrain the emission mechanism. In
fact, the spectral variations observed during long, high signal-to-noise
observations of NGC~5548 and NGC~7469 (Petrucci et al. 2000, Nandra et al.
2000) have provided strong support to the Comptonization hypothesis as the
origin of X-rays in AGN. In the former case, during a long {\it Beppo} SAX
observation the spectrum became ``harder" (i.e. flatter) and the corona
temperature decreased together with the $[2-10]$ keV flux whereas the
total luminosity remained essentially constant. In NGC~7469, simultaneous
{\it RXTE} and {\it IUE} observations revealed strong X-ray spectral
changes which were correlated with the UV flux. In both cases, the
observed spectral variations can be naturally explained in terms of
thermal Comptonization models since an increase in the UV photons reduces
the temperature of the hot corona and produces a ``softer" (i.e. steeper)
X-ray power law slope.

In this work we present the results from a spectral variability analysis
of four Seyfert galaxies, namely NGC~4051, MCG~-6-30-15, NGC~5506 and
NGC~5548. These objects have been monitored with {\it RXTE} on a regular
basis since 1996. The main purpose of the observations was to compute
accurately the shape of the power density spectrum over a broad range of
time scales in order to reveal the existence of characteristic time scales
(low or high frequency ``breaks").  These time scales can in principle
provide model-independent black hole mass estimates through their
comparison with similar time scales that have long been identified in the
power spectra of Galactic X-ray binaries. The results of this study are
presented elsewhere (Uttley, M$^{\rm c}$Hardy \& Papadakis 2001).  Here we
focus on the variability properties of the energy spectra of the sources.

All the sources are strongly variable in the X-rays. MCG~-6-30-15 and
NGC~5548 are classified as Seyfert 1 nuclei, NGC~5506 is a Seyfert 2
galaxy while NGC~4051 is a Narrow Line Seyfert 1 (NLS1) galaxy. They are
typical examples of their class and taken as a whole they can be
considered as a representative group of the nearby Seyfert galaxies. 

There are two main advantages of the {\it RXTE} light curves that we
study in the present work: {\it i)} they are of high quality (i.e. high
signal to noise ratio) and {\it ii)} they span regularly a long time
interval ($\sim 3$ years for each source).  As a result, we are able to
study spectral variations covering the highest, lowest and intermediate
flux states over a few years.  This is not possible to achieve with the
typical X-ray observations of AGN which last for $\sim$ few days
(maximum).

Using the same {\it RXTE} data, Lamer, M$^{\rm c}$Hardy \& Uttley (2000)
and Lamer et al. (2001) have already presented a study of the spectral
variability of NGC 5506 and NGC 4051. Their study was based on model
fitting of flux-averaged energy spectra. In this paper, we follow a
different approach.  Typically, each source was observed for $\sim 1$ ksec
during each pointing.  Although this is not long enough for a direct study
of the energy spectrum it allowed us an accurate flux measurement in the
$[3-15]$ keV band. Using the {\it RXTE} light curves at different energy
bands we computed hardness ratios, produced color-flux diagrams and used
these diagrams to investigate the properties of spectral variations.  As a
result we are able to study the spectral variations in a systematic way,
i.e. at each individual observation without considering broad ``flux
states" for each source. In this way, we cannot investigate whether a
given model describes well the energy spectrum of the sources. However,
using the simple ``power law + iron line + reflection component" model
which provides a good representation of the X-ray spectrum of AGN, we were
able to study in detail the relationship between the continuum, line flux
and reflection component.

The paper is organized as follows. In the following section we present our
data and in Section 3 we present the results from the analysis of the
variability of the hardness ratios.  In Section 4 we use the standard
``power law plus reflection component plus narrow iron line" model to
compare quantitatively the observed trends in the color-flux diagrams with
specific hypothesis on the variability properties. In Section 5 we compare
our results with previous studies, and in Sections 6 and 7 we present the
discussion of our results and our conclusions.

\section{Observations and Data reduction}

We use data obtained with the PCA on board {\it RXTE} covering a three
year period during observing cycles 1-3 when the PCA gain setting was
constant.  Details of the data sampling scheme and of the time of the
first and last observation of the sources can be found in Uttley et al.
(2002). Briefly, the four sources were observed $\sim 120$ times during
the period from April/May 1996 until December 1998/February 1999.  Each
observation was $\sim 1$ ksec long and the observing scheme was designed
so that the broadest range of time scales would be covered with the
minimum observing time. The objects were observed once every week in the
first year of the monitoring campaign and then every two weeks. During
some periods the monitoring was more intensive (twice daily or daily for a
period of two up to four weeks) to sample the shorter time scale
variations. The data were reduced using FTOOLS V4.2. We used data from the
top layer of the PCUs 0,1,2 only and estimated the background light curves
using the $L7$ background model (for details see Uttley et al. 2002).

The background subtracted $[2-10]$ keV light curves of the four sources,
normalized to their mean, are shown in Figure 1. All are significantly
variable. The peak to peak variability amplitude is $\sim 40, 5, 3 $ and
$4$ for NGC~4051, MCG~-6-30-15, NGC~5506 and NGC~5548 respectively.
Therefore NGC~4051 stands out as the most variable source as is typical
of NLS1 nuclei (e.g. Leighly 1999a). During the {\it RXTE} observations,
NGC~4051 went through an unusually ``low-state" when the source reached
a very low flux level (the period between Day 600 and 750 in Figure 1).
In fact, there is the possibility that the central source even
``switched-off", leaving only the reflection spectrum from cold matter
to be detected (Guainazzi et al. 1998, Uttley et al. 1999). For that
reason, we excluded the points in the NGC~4051 light curves between Day
600 and 750 from the data analysis.

Finally, the {\it RXTE} observations of NGC 5548 are contaminated by the
nearby BL Lac object 1E 1415.6+2557. Chiang et al. (2000) have estimated
that the contaminating contribution of this object to the $2-10$ keV NGC
5548 light curves (using the same three PCUs as ours) is $\sim 2$
counts/sec with a modulation smaller than $0.8$ counts/sec. Since we
observe a peak to peak count rate difference of $\sim 20$ counts/sec, the
1E 1415.6+2557 contribution to the overall variability properties of
NGC~5548 should not be crucial.

\section{Data Analysis}

\subsection{Hardness Ratios}

We extracted light curves in the following energy bands: $[3-5]$, $[5-7]$,
$[7-10]$ and $[10-15]$ keV. These intervals were chosen so as to separate
as much as possible the various emission components in the spectrum of the
objects. The first and third band should be representative of the primary
continuum mainly, while the iron line and the reflection component
emission should contribute in the $[5-7]$ keV and $[10-15]$ keV bands
respectively.  From these light curves we calculated three hardness
ratios: $HR1=C_{7-10keV}/C_{3-5keV}$, $HR2=C_{7-10keV}/C_{5-7keV}$ and
$HR3=C_{10-15keV}/C_{7-10keV}$ ($C_{E1-E2}$ represents the count rate in
the energy band $E1-E2$). These are sensitive to variations : i) of the
continuum shape ($HR1$), ii) of the continuum to the iron line ratio
($HR2$) and iii) of the reflection component to the continuum ratio
($HR3$).

Figure 2 shows a plot of the hardness ratios as a function of time since
the beginning of the $RXTE$ observations for each source. The solid lines
show the $HR$ weighted average values ($\bar{HR}$). These values, together
with the root mean square variability amplitude (i.e. $\sigma_{rms}$) of
the $HR$ light curves are reported in Table 1. A constancy $\chi^2$ test
shows that all the $HR$ values exhibit statistically significant
variations, except from the $HR3$ plot of NGC~5548 for which the
significance is only marginal. The $HR1$ light curve of NGC~4051 has the
largest $\sigma_{rms}$ among the $HR1$ light curves.  This source shows
the largest amplitude flux {\it and} spectral variations. 

The main conclusion from Table 1 is that the mean $HR$ values are roughly
similar between the different objects. This implies that the mean energy
spectrum has approximately the same spectral shape in all sources.
However, there are also some differences. For example, NGC~4051 shows the
smallest $\bar{HR}$ values, indicating that its energy spectrum is softer
than the spectrum of the other sources. This is consistent with results
from previous studies which have showed that the X-ray energy spectra of
NLS1s are ``softer" than the spectra of typical S1 galaxies (e.g. Leighly
1999b). The $\bar{HR}$ values of NGC~5506 (the only S2 source in our
sample) are similar to the respective values of NGC~5548 and MCG~-6-3015
(the S1 sources in the sample) except from $\bar{HR1}$.  Its large value
is probably due to the presence of a large absorbing column towards that
source which causes the energy spectrum to appear ``harder" than the
spectrum of other sources.

\subsection{Color-Flux diagrams}

Having established the existence of significant spectral variability in
all sources, we then investigated whether the spectral variations are
correlated with the source flux.  Figure 3 shows the plot of $HR1$ as a
function of the normalized $([3-5]+[7-10])$ keV count rate
($C_{(3-5keV+7-10keV)_{norm}}$). Figures 4 and 5 show a plot of $HR2$ and
$HR3$ as a function of the normalized $[3-5]$ keV count rate
($C_{(3-5kev)_{norm}}$). In Figure 3, we have chosen the $([3-5]+[7-10])$
keV count rate as representative of the continuum flux, since the sum of
the source signal in the two bands should minimize the possibility to
introduce artificial correlations due to the inter-dependency of the $HR1$
and $C_{(3-5+7-10keV)}$ variables. In Figures 4 and 5, we have used only
the $[3-5]$ keV count rate since both $HR2$ and $HR3$ are independent of
$C_{(3-5kev)}$. In all cases, we have normalized the count rates to their
mean in order to compare the color-flux plots of the four sources.

The points in the color-flux plots of all the sources cluster in a well
defined, ``continuous" region rather than forming a scatter diagram or
filling separate ``islands". Although it is not clear from the power
density spectra of the sources whether we have observed their
maximum/minimum flux states with the present {\it RXTE} observations
(see discussion in Uttley et al. 2002) we are certain that we have
observed all the ``intermediate" flux states of each source between
their largest and lowest flux state in the $RXTE$ light curves.

The color-flux trends are similar for each object: $HR1$
systematically decreases (i.e. the energy spectrum softens) as the flux
increases. A similar trend is observed for $HR3$. On the other hand, the
$HR2$ variations appear to be independent of the source flux.

In order to quantify the above relationships we fitted the data for each
galaxy using a power law function of the form: $HR=a_{1}C_{norm}^{b_{1}}$.
We also tried a linear relationship (i.e. $HR=a_1+b_1\times C_{norm}$) but
the goodness of fit was worse in all diagrams. The model fitting was done
taking account of the errors in both variables. The best fitting results
are listed in Table 2. The best fitting $b_{1}$ values in the
$HR2/C_{{(3-5keV)}_{norm}}$ diagrams are consistent with zero (contrary to
the $HR1/C_{{(3-5keV+7-10keV)}_{norm}}$ and $HR3/C_{{(3-5keV)}_{norm}}$
diagrams). This result shows that the $HR2$ variations are indeed
independent of the source flux. The only exception is NGC 4051 which shows
a $HR2/C_{{(3-5keV)}_{norm}}$ slope which, although small, is
significantly different from zero.

NGC 4051 also shows the largest absolute $b_{1}$ values in all diagrams.
The spectral variations in this source are thus a ``steeper"  function
of the flux than in the other sources, i.e. for the same amplitude flux
variations, the spectral variations in NGC~4051 are larger. Furthermore,
the NGC~5506 $b_{1}$ value in the $HR1/C_{{(3-5keV+7-10keV)}_{norm}}$
diagram is different from the values of NGC~5548 and MCG~-6-30-15.  This
difference cannot be attributed to the presence of a large absorbing
column towards NGC~5506, unless the column density varies with the
X--ray source flux. However such behavior has never been observed in S2
nuclei (Risaliti, Elvis and Nicastro 2001).

As Figures 4,5 and 6 show, a power law describes well the overall trend in
all the color-flux plots. However, it does not give give a statistically
acceptable fit, except for the $HR3/C_{(3-5keV)_{norm}}$ diagrams (this
is probably due to the larger errors associated with the HR3 values).
There is significant scatter around the mean trends which shows that the
spectra are significantly variable at any given ``flux state" of the
sources as well.

The spectral variability behavior of the sources could be studied
with the frequently used color-color diagrams as well. For example, $HR2$
vs $HR1$ diagrams could be used to investigate, directly, whether the
variations of the line's $EW$ are correlated with the primary spectrums
slope changes. However, as we show in the Appendix, the fact that both
$HR1$ and $HR2$ use the count rate in the $[7-10]$ keV band, implies that
there could exist misleading correlations in the $HR2/HR1$ diagram. For
that reason, we do not use any color-color diagrams to investigate the
spectral variability behavior of the sources.

\section{Determination of spectral parameters from the
 hardness ratios}

In order to understand the spectral variability behavior of the four
sources in terms of commonly used spectral parameters, we used the widely
used model consisting of a power law plus Fe line plus a reflection
component. This model fits well the average X-ray/$\gamma$-ray spectrum of
Seyfert 1 nuclei, especially in the 2--20 keV energy range (Nandra \&
Pounds 1994). We computed the model hardness ratios under different
assumptions and compared them with the data as described in detail below.
This comparison allows us to derive the main spectral characteristics
(i.e. spectral index, reflection and iron line equivalent width) and to
discuss the spectral variability exhibited by the different objects in our
sample.

\subsection{The method}

The model consists of a power law continuum of the form: $f_{E}=N
E^{-\Gamma}$, where $N$ is the normalization at 1 keV.  We included a
Compton reflection component using the PEXRAV model (Magdziarz \&
Zdziarski 1995) in XSPEC. This extra component calculates the X-ray
spectrum when a source of X-rays is incident on optically thick, neutral
(except hydrogen and helium) material.  We fixed the inclination angle
for the disk at $i=30^{\circ}$ (the shape of the reflection spectrum
below 20 keV is relatively independent of the inclination angle). The
iron and light element abundances were kept fixed at the solar abundance
values. The strength of the reflection component is governed by the
parameter $R$, representing the strength of the reflected signal
relative to the level of the incident power-law continuum. We also added
a narrow Gaussian line ($\sigma_{line}=0.1$ keV) to simulate the iron
line. The line energy was kept fixed at $6.4$ keV so that the line is
completely characterized by its equivalent width ($EW$). Finally, except
from NGC~5506, we do not consider the effects from neutral absorption in
our calculations, since the absorbing column towards the direction of
the sources is consistent with the Galactic value and does not affect
the X-rays at energies $> 2$ keV. For NGC~5506 we fix the value of
$N_{H}$ at $3.6\times 10^{22}$ cm$^{-2}$ (Perola 1998).

Assuming that this model gives a good representation of the spectral shape
of the sources, we have a one-to-one correspondence between the data (the
count rate in the four energy bands) and the model parameters ($N$,
$\Gamma$, $EW$ and $R$). As a result, the observed count rates can be used
to calculate the model parameters as described below.

Using XSPEC (v11.0), we computed the model count rate expected in
different energy bands from the power law continuum, the reflection
component and the Iron line for different values of the spectral index
$\Gamma$, but fixing $N=1$, $R=1$, and $EW=250$ eV
($C_{E1-E2}^{pl}(\Gamma), C_{E1-E2}^{R}(\Gamma)$ and
$C_{E1-E2}^{Line}(\Gamma)$, respectively, where $(E1-E2)=[3-5], [5-7],
[7-10]$ and $[10-15]$ keV). Then, for each observation $i$, the observed
count rate at each energy band is given by,

\begin{equation}
C_{E1-E2}=N_i\left[C_{E1-E2}^{pl}(\Gamma_i)+R_iC_{E1-E2}^{R}(\Gamma_i)+
\frac{EW_i}{250}C_{E1-E2}^{line}(\Gamma_i)\right],
\end{equation}

We have thus a system of four equations with four unknown variables
($N_i,\ \Gamma_i,\ R_i$ and $EW_i$) which is easily solvable for each
pointing.

Since we are interested in the overall spectral behavior of the four
sources during the total observational period, we did not solve the system
of equations (1) for each observation.  Instead, we used these equations
to calculate the spectral slope and normalization at various flux states
along the best fitting lines shown in Figure 3.  Then we computed the
$HR2$ and $HR3$ model ratios considering two different possibilities for
the response of the reflecting material to the continuum variations: (a)
The material responds with a short delay to the continuum variability, in
which case the line's $EW$ and the reflection component's parameter $R$
should remain constant during the spectral variations. This is expected in
the case when the material is located close to the X-ray source (i.e. the
accretion disk). (b) The reflecting/reprocessing material does not respond
simultaneously to the fast, intrinsic variations, as expected if the
reprocessor is located away from the central source (i.e. the obscuring
torus). In this case the line and the reflection component {\it flux}
should remain constant.

The model $HR2/C_{(3-5keV)_{norm}}$ and $HR3/C_{(3-5keV)_{norm}}$ curves,
in the case of constant $EW$ and $R$, are plotted in Figures 4 and 5. In
Figure 4, the dashed, dotted and dashed/dotted lines show the model curves
in the case of constant $EW=0, 250$ and $500$ eV respectively ($R$ is kept
constant to 1). The dashed, dotted and dashed/dotted lines in Figure 5
show the model curves in the case of constant $R=2,1$ and 0 respectively.

To produce the model diagrams in the case of a remote reflector, we used
the average $HR$ values, listed in Table 1, to find the spectral
parameters (i.e., $\Gamma$, $N$, $EW$ and $R$) of the time average
continuum of the four sources. Using XSPEC, we then deduced the flux of
the line and of the reflection component in the average spectra. These
fluxes were kept constant in order to estimate the model $HR2$ and $HR3$
values, at each $C_{(3-5 keV)_{norm}}$, using the respective ($\Gamma,N$)
values (computed for $R=0$). The corresponding model curves are also
plotted in Figures 4 and 5 (solid lines).

\subsection{Results}

\subsubsection{The average spectra}

Using equation (1) and the mean $\bar{HR1}$, $\bar{HR2}$ and $\bar{HR3}$
values of each object, we estimated the mean spectral index
$\langle\Gamma\rangle$, mean reflection normalization $\langle R \rangle$
and mean iron line equivalent width $\langle EW\rangle$. The results are
listed in Table 3. The values in this table show that NGC~4051 possesses
the largest average $R$ and $EW$ values among the four sources.  The mean
spectral characteristics of the Seyfert 2 NGC~5506 are similar to those
found for the two S1 objects, NGC~5548 and MCG~-6-30-15.

Our results are in good agreement with the respective values that have
been reported in the past, based on model fitting of the full band
energy spectra. For example, Lamer et al. (2000, 2001), find an average
value of $\Gamma\sim 2.1, R\sim 1.2$ and $EW\sim 300$ eV for NGC~5506,
and $\Gamma\sim 2.05$, $EW\sim 400$ eV for NGC 4051. These values are
consistent with the values listed in Table 3 for the same objects.  The
average values for the line's $EW$ that have been reported in the past
for MCG~-6-30-15 and NGC~5548 (Wang et al. 1999, Lee et al. 1999, and
Chiang et al. 2000) are in agreement with the respective values in Table
3. The average $R$ value that we find for MCG~-6-30-15 also agrees with
the values that Lee et al (1999) report, while it is slightly larger in
the case on NGC~5548 (0.8 as opposed to $\sim 0.5$ in Chiang et al
2000). Finally, the slope values that we find for MCG~-6-30-15 and
NGC~5548 are in good agreement with the values reported in the past for
the same objects.

Interestingly, we find a correlation between $\langle R \rangle$ and
$\langle\Gamma\rangle$ as well as a correlation between $\langle
EW\rangle$ and $\langle\Gamma\rangle$. When comparing the average
spectra of the four sources, we see that the sources which show harder
spectra show smaller (average) $R$ and $EW$ values as well. This
correlation is in agreement with the claims of Zdziarski et al. (1999).

\subsubsection{The primary continuum variability}

The dashed lines in Figure 3 show the best fitting curves to the observed
color-flux diagram of the four sources. On these lines, we have plotted
the spectral slope values for a few normalized count rate values. These
slope values are computed as explained in Section 4.1 in the case of
$R=1$. According to the model predictions, the observed $HR1$ variations
correspond to significant continuum slope changes.  Differences in the
spectral index up to a $\Delta \Gamma \sim 0.3, \sim 0.2$ and $\sim 0.2$
are necessary to explain the $HR1$ variability in MCG~-6-30-15, NGC~5506
and NGC~5548 respectively. In NGC~4051, differences larger than $\Delta
\Gamma \sim 1$ are needed to explain the observed $HR1$ variations.
Although we have excluded from our analysis the points that correspond to
the ``off-state" of this source, Uttley et al. (1999) have shown that the
decline of the source towards this ``off-state" is gradual. If the
spectrum in this state is simply a reflection component produced by the
torus and remains present at all times, then this is probably the dominant
component in the X-ray spectrum of NGC 4051 in the observations that have
the lowest count rate in Figure 3. As a result, the spectral slope that we
measure could be affected by the shape of the reflection component, and
the primary slope could be steeper than what we estimate. In fact, Lamer
et al. (2001), find slope variations between $\Gamma\sim 1.6$ and $2.3$
when they subtract from the source's spectra at the various flux states
the ``off-state" spectrum. As expected, their upper limit is consistent
with ours ($\sim 2.4$, see Figure 3) while their lower limit is larger
than ours.

In order to investigate further the effects of a constant flux reflection
component on the spectral slope variability, we used the flux of the
reflection component in the average spectrum of the sources (see Section
4.1) and assumed that it remains constant. In this case, since the
reflection component contributes to the [$7-10$] keV band more than it does
to the [$3-5$] keV band, it can result in an artificial flattening of the
photon index, mainly when the source is at low flux states. For that
reason, taking into account the constant reflection component flux at each
energy band, we computed again the [$N, \Gamma$] values (using the best
fitting line to the $HR1$ vs flux plot, as explained in Section 4.1).
Figure 6 shows again the plot of $HR1$ as a function of the normalized
([$3-5$] + [$7-10$]) keV count rate, with the $HR1$ values of NGC~4051
during the ``low-state" included as well (in order to examine the
intrinsic spectral slope values when the source is dominated by the
reflection component). On the best fitting lines, we mark the new spectral
slope values for a few normalized count rates, like we did in Figure 3.

As expected, due to the presence of a constant flux reflection component,
smaller $\Delta \Gamma$ variations are necessary in order to account for
the observed $HR1$ variations, and the ``boundary'' $\Gamma$ values are
now steeper. However, the differences between the results shown in Figures
3 and 6 are small. The assumption of a constant flux reflection component
alone cannot account for the observed $HR1$ variations. Intrinsic spectral
slope variations of the order of $\sim 0.2$ are still needed in order to
explain the spectral variations in MCG~-6-30-15, NGC~5506 and NGC~5548.
Even in NGC~4051, intrinsic variations of $\Gamma$ between $\sim 1.5$ and
$\sim 2.4$ are still implied by the $HR1$ variations. In fact, despite the
presence of the constant flux reflection component, a primary component
(albeit with a small normalization) with a very flat, variable slope needs
to be present in order to account for the $HR1$ values when the source is
in the low/off-state.

More important effects on the implied $\Gamma$ variations can have the
variability behavior of the broad red wing of the Fe line in MCG~-6-30-15.
Based on $ASCA$ observations, Iwasawa et al. (1996)  found that the iron
line in this source consists of a narrow core at an energy of $\sim 6.4$
keV, and a broad wing extending to below 5 keV. While the narrow component
cannot contribute significantly to the count rate in either the [$3-5$] or
[$7-10$] keV band, the broad red wing will contribute to the observed
[$3-5$] keV count rate. Iwasawa et al. (1996) found that the broad
component can be fitted well with a Gaussian centered at $5.5$ keV with a
dispersion of $0.64$ keV. They also found that the $EW$ of this component
component is variable, decreasing from $\sim 600$ eV to $\sim 100$ eV with
increasing flux. Consequently, when the source is at low state, the broad
wing can contribute significantly to the [$3-5$] keV band count rate,
resulting in a flatter intrinsic slope.

In order to investigate this possibility, we recalculated the [$N,
\Gamma$] values for MCG~-6-30-15, assuming that the $EW$ of the line's red
wing decreases from $\sim 0.6$ keV to $\sim 0.1$ keV, when the source is
at the lowest/highest states in the present $RXTE$ light curve. As
expected, when the source is above its average flux state, the presence of
the broad iron line wing does not affect significantly the estimation of
the spectral slope value. However, as the flux decreases, the slope values
that we estimate become flatter. At the lowest flux state (i.e. when
$C_{{(3-5keV+7-10keV)}_{norm}}\sim 0.4$ in the MCG~-6-30-15 plot of
Figures 3, 6) we find a value of $\Gamma\sim 1.7$, as compared to
$\Gamma\sim 1.9$ that we find when we do not consider the effects of the
line's red wing. We conclude that, if the variability behavior of the
extended red wing of the Fe line in MCG~-6-30-15 on long time scales is
similar to the behavior exhibited during the $ASCA$ observations, the
observed $HR1$ variations of this source imply variations of the primary
spectral slope of the order of $\Delta\Gamma\sim 0.4 - 0.5$.

Finally, using the [$N, \Gamma$] values at each flux state as shown in
Figure 3, we computed the primary power law flux in the energy range
between $1-300$ keV. Figure 7 shows a plot of this flux as function of the
normalized source count rate. In NGC 4051, we find that the luminosity
variations have an amplitude much smaller than that of the observed flux
variations. In this source, the large amplitude $[3-15]$ keV flux
variations appear to be caused mainly by the large amplitude continuum
slope variations. In the other sources, the peak to peak amplitude of the
luminosity change is $\sim 2$. This is smaller but comparable with the
amplitude of the observed flux variations. However, since the variability
behavior of the high energy cut--off component cannot be studied with the
present {\it RXTE} data, it is possible that the amplitude of the
luminosity variations is smaller in reality if there are significant cut
off variations. In fact, Petrucci et al. (2000) find that in NGC~5548 (the
source that shows the largest amplitude luminosity variations in Figure 7)
the cut--off variations are associated with the spectral slope variations
in such a way so that the total X-ray luminosity remained constant during
the long {\it Beppo} SAX observation.

\subsubsection{The line variability}

Filled squares in Figure 4 show the average $HR2$ values over various
flux (i.e. $C_{(3-5keV)_{norm}}$) bins. The size of the bins was
variable so that 15 points were included in each one. These points
indicate the average $HR2$ behavior as a function of the soft band flux
and are close to the best fitting lines of the
$HR2/C_{{(3-5keV)}_{norm}}$ diagrams.

The ``constant $EW$" model predictions do not agree well with the
average $HR2$ values. Instead, the model curves suggest rather complex
variations for the line's $EW$ in all sources. It appears that, on
average, the line's $EW$ decreases as the source flux increases. For
example, for NGC 5548, when the source is at the lowest flux state the
line's $EW$ is $\sim 250$ eV. As the source flux increases, the line's
$EW$ decreases to $\sim 50$ eV. In NGC 4051, when the normalized soft
band count rate is below $\sim 0.7$, the line's $EW$ appears to increase
strongly, reaching a value larger than $500$ eV at the very low flux
states. The tendency of the line's $EW$ to decrease with increasing
source flux also appears in the color-flux diagrams of NGC 5506 and
MCG~-6-30-15 but the amplitude of the $EW$ variations is smaller in
these two sources.

In the same Figure, we have also plotted the ``constant line flux" model
curves. In this case, the model predictions are consistent with the
average $HR2$ values, except perhaps NGC 5506 where the model curve lies
above the average $HR2$ values. We can get an agreement between the model
predictions and the NGC 5506 data if we increase slightly the flux of the
line that we add to all the model spectra.

Column 2 and 3 in Table 4 lists the slope of the model curves shown in
Figure 4. The $b_{CEW}$ values correspond to the slope of the model curves
under the assumption of constant line's $EW$, while $b_{CLF}$ is the slope
of the constant line flux model curves. Comparison of these values with
the best fitting slope values listed in Table 2, shows clearly that the
constant flux model predictions agree better with the data; their slopes
are consistent with the best fitting slope values. On the other hand, the
difference between the constant $EW$ model curve slopes and the best
fitting slopes is larger than $3\sigma$ in all sources.

\subsubsection{The reflection component variations}

As before with Figure 4, filled squares in Figure 5 show the average $HR3$
values as a function of the soft band flux. In the same Figure, we also
show both the ``constant $R$" and ``constant reflection flux" model
curves. 

Due to the large scatter in the $HR3/C_{{(3-5keV)}_{norm}}$ diagrams of
all sources, and the similarity of the model curves under the two
different assumptions, it is not clear which model curves fit better the
data. In the MCG~-6-30-15 and NGC 5506 plots, comparison between the
constant $R$ model curves and the average $HR3$ values show that $R$
slightly decreases with increasing source flux, in agreement with the $EW$
dependency on the flux in these sources. However, most of the average
$HR3$ points are close to the constant $R=1$ model curve, and only the
lowest flux $HR3$ value is clearly closer to the constant $R=2$ curve. In
NGC 5548, $R$ appears to remain constant during the spectral variations at
a value smaller than 1. Similarly, in NGC 4051, the average $HR3$ values
imply spectral variations under constant $R$ (with a value between 1 and
2). Only the lowest flux points suggest a decreasing $R$ value with
decreasing source flux.  However, this is again caused by the fact that
the spectral slope that we measure from observations with the lowest flux
represents the shape of the reflection component itself. Hence the
$\Gamma$ values are larger and the $R$ values are smaller than in reality
(see discussion in section 4.2.1)

The constant reflection flux model curves also provide a good fit to the
average $HR3$ values in all sources. Columns 4 and 5 in Table 4 list the
slopes of the constant $R$ and constant reflection flux model curves
($b_{CR}$ and $b_{CRF}$ respectively). Comparison between these values and
the best fitting slopes of the $HR3/C_{{(3-5keV)}_{norm}}$ diagrams
(listed in table 2) shows that all model curves are consistent with the
data. The only exception is the $b_{CR}$ value of NGC 5506, which show a
$5\sigma$ difference with the best fitting slope.

\section{Comparison with previous studies}

The spectral variability properties of the four sources that we study in
the present work have already been addressed previously. In most cases,
the results are based on spectral fitting of the power law plus reflection
model to the observed energy spectra.

All studies have shown that as the source brightens the continuum
spectrum steepens. Differences in the spectral slope up to $\Delta
\Gamma \sim 0.3, 0.2$ and $0.15$ have been observed in MCG~-6-30-15 (Lee
et al. 1999), in NGC~5506 (Lamer et al. 2000)  and in NGC~5548 (Chiang
et al. 2000; Petrucci et al. 2000).  Our results are consistent with the
previous studies. We find that the observed $HR1$ variations correspond
to larger spectral slope changes. This is because we have detected flux
variations of a larger amplitude. In the case of NGC~4051 we find a
larger $\Delta \Gamma$ variations compared to Lamer et al. (2001), but
as we discussed in section 4.2.1, this is due to the significant
contribution of the ``off-state" spectrum to the source's spectra at low
flux states.

Variability of the iron line's $EW$ and of the reflection fraction $R$
have also been observed. Using ASCA data of 39 AGNs, Nandra et al. (1997)
found a clear decrease in the iron line's $EW$ with increasing source
luminosity. Lee et al. (2000), Lamer et al. (2000) and Chiang et al.
(2000) also find that the iron line's $EW$ decreases with increasing
source flux in MCG~-6-30-15, NGC~5506 and NGC~5548 respectively. These
results are consistent with ours. Furthermore, the dependence of the
combined ``narrow" and ``disk" line's $EW$ on the source's flux level that
Lamer et al. (2001) find in NGC~4051 is entirely consistent with our
results shown in Figure 4. On the other hand, Wang et al. (1999) have
observed a positive correlation between the line's $EW$ and the source
flux in NGC~4051 using ASCA data, contrary to our results. We note however
that NGC 4051 was ``weakly'' variable (factor of 2) during the ASCA
pointing in comparison to variations of more than a factor 10 in our case.
Furthermore our $HR2/C_{{(3-5keV)}_{norm}}$ diagram for this object shows
large scatter around the mean trend. We thus believe that the results of
Wang et al. (1999) are consistent with our conclusions.

Lamer et al. (2000) find that the reflection fraction in NGC~5506
decreases with source flux; in particular it is not correlated to the
continuum slope as derived from direct fitting of {\it RXTE} spectra.
Chiang et al. (2000) find that the reflection fraction remains constant in
NGC~5548. These results are also consistent with ours. On the other hand,
Lee et al (2000) find that the reflection fraction in MCG~-6-30-15
increases with source flux (and is anticorrelated with the line's $EW$)
contrary to what we obtain. Finally, in NGC 4051, Lamer et al. (2001) find
that the reflection fraction of the ``disk component" (i.e. the primary
spectrum without the contribution of the ``off-state" spectrum) is very
small. Therefore if the only reflection component is the one that is
observed in the ``off-state" then its flux should remain constant. As
Figure 5 shows, our results are entirely consistent with this possibility.

\section{Discussion}

We have studied in detail the spectral variability of four Seyfert
galaxies (NGC~4051, MCG~-6-30-15, NGC~5506 and NGC~5548) over a $\sim 3$
years period. Using $RXTE$ data we computed hardness ratios which are
sensitive to the primary slope, iron line and reflection component
variations ($HR1, HR2$ and $HR3$ respectively).  Our results show that all
the sources exhibit significant spectral variability. We find that $HR1$
and $HR3$ decrease as the source flux increases while the $HR2$ variations
are independent of the source flux. The spectral variability behavior of
the Narrow Line Seyfert 1 galaxy NGC 4051 and the Seyfert 2 galaxy NGC
5506 is similar to the behavior of MCG~-6-30-15 and NGC 5548, the two
Seyfert 1 nuclei in our sample. There are also minor differences.  NGC
4051 in particular shows the largest spectral slope variability amplitude
among the four sources.

Our approach to study the spectral variability of the sources is different
from previous studies. Instead of a detailed spectral study that uses the
full energy spectrum (usually in the $\sim 2-15$ keV band with {\it RXTE}
data) at a few epochs (at best), we use a large number of short exposures
over a long period, and compare the observed color-flux diagrams with the
predictions of the simple ``power law plus Compton reflection component
plus iron line" model. The hardness ratios cannot be used in order to
investigate whether the model can fit well the X-ray spectrum of the
sources or not, however, it is known from past studies that this model does
provides a good representation of the X-ray energy spectra of AGN. In this
case, due to the large number of observations which are regularly
distributed over a long period (much longer than the typical duration of
the AGN observations) the hardness ratios can be used to investigate,
accurately, which components in the X-ray spectra of AGN are responsible
for the correlations that we find in the color-flux diagrams.

We did this by constructing various model color-flux curves. For the
$HR2/C_{(3-5kev)_{norm}}$ and $HR3/C_{(3-5kev)_{norm}}$ model curves we
considered two possibilities for the reflecting and the line producing
material. First, we produced model curves assuming that the material
responds fast to the continuum variations. In this case the line's $EW$
and the reflection parameter $R$ remain constant during the spectral slope
variations. Alternatively, in the case when the material does not respond
quickly to the continuum variations, it is the flux of the line and the
reflection component that remains constant.

Our main results from the comparison of the observed color-flux plots with
the predictions of the ``power law + reflection + iron line" model are the
following:

1) The primary continuum slope is variable. Averaging HR1 in fixed flux
intervals, the derived slope is positively correlated with the soft band
flux. At each flux state, there are also significant slope variations
around the mean slope that corresponds to that state.

2) As the soft band flux increases, the line's $EW$ decreases (Figure
4). There is an indication that the reflection parameter $R$ also
decreases with increasing flux in the case of NGC 5506 and MCG~-6-30-15
(Figure 5). In NGC 4051 and NGC 5548, $R$ remains constant during the
primary spectral and flux variations. In any case, the $HR3$ color-flux
diagrams do not show any increase of $R$ with increasing source flux
(i.e. as the primary spectrum steepens).

\subsection{Power-law slope variability}

Spectral slope variability is predicted by thermal Comptonization models.
An increase of the soft input photon flux will result in a reduction of
the corona temperature, a steepening of the intrinsic spectrum and hence
an increase in the soft X-ray band flux. This response of the X-ray
spectrum to the soft input photon changes has already been observed in
NGC~7469 (Nandra et al. 2000) and NGC~5548 (Petrucci et al. 2000). In the
simplest case, we expect a correlation between the spectral slope and the
soft band flux, as observed in Figure 3. It is remarkable how similar the
average spectral behavior is for three of the sources and for all the
intensity states covered. In particular 5506, a Seyfert 2, follows the
same behavior as 5548 and MCG~-6-30-15.

The fact that the spectral slope variations of NGC 4051 have a larger
amplitude could be due in part to a larger variation of the strong soft
component known to be present in this source, affecting mainly the $[3-5]$
keV band, hence resulting in a steeper spectrum. As we already discussed
in Section (4.2.1), the very flat spectra that we observe at the lowest
flux states could be the result of the fact that in these cases the
reflection component is prominent in the energy spectrum (affecting mainly
the $[7-10]$ keV band) so that the estimated slope determines mainly the
shape of this component instead of the primary continuum.

Apart from the average trend, the $HR1-$flux plots also show significant
scatter around the best fitting lines. Therefore, spectral slope
variations happen at constant flux as well. Since different slope values
result in the same soft band flux, these variations should be associated
with luminosity changes. Apart from variability induced by the
Compton-cooling process, other processes that affect the corona properties
(like and the corona heating process for example) should also be variable.

\subsection{Iron line and reflection component variability}

In NGC 5506 and MCG~-6-30-15, the line's $EW$ and the reflection
parameter $R$ vary proportional to each other (i.e. in the same way with
the source flux). This implies that the material which emits the iron
line is also responsible for the reflection hump. In the other two
sources, $R$ remains constant while $EW$ decreases with the source flux.
If this is the case, the line emitting and reflecting material should
not be the same. However, it would be difficult to explain why matter
that produces the iron line cannot produce the reflection hump as well
(and vice versa).

Changes in the line's $EW$ and the reflection parameter $R$ could be the
result of variations in the geometry of the system. For example, if the
solid angle subtended by the reflector decreases, both the line's $EW$ and
the reflection parameter should also decrease. A relativistic outflow of
the X-ray source could produce such an effect. However, at the same time,
it should also be able to produce a softening of the primary spectrum and
an increase in the total X-ray luminosity in order to explain the
simultaneous increase of the soft band flux as both $EW$ and $R$ decrease.

The observed $EW$ and $R$ variability behavior of the sources can be
explained in a consistent way if the reprocessing material does not
respond instantly to the continuum spectral variations. As the solid lines
in Figures 4 and 5 show, the assumption of constant flux for the line and
the reflection component during the spectral variations can account for
the mean trends in the $HR2$ and $HR3$ color-flux diagrams. As the source
luminosity and hence the continuum normalization increases, the soft band
flux increases (see Figure 7) while both $EW$ and $R$ decrease because the
flux of the line and reflection component remain constant. Since the
average spectral properties of the sources are not identical (see Table
3), the constant flux model curves can explain quite well the spectral
variations although the best fitting lines in the color-flux plots are not
the same for all sources. Even in the cases where $R$ remains constant
during the flux variations (e.g. NGC 5548) the model curves can still fit
well the color-flux diagrams. The reason is that the constant flux and
constant $R$ model curves are quite similar; we need data with smaller
uncertainty to discriminate between the two cases.

A natural explanation for the fact that the reprocessor does not respond
to the fast intrinsic variations is that it is located away from the
central source, i.e. it is associated with the obscuring torus in the
objects. Interestingly, this possibility, at least in the case of
NGC~5548, is supported by a recent $CHANDRA$ observation of this source
which has revealed the existence of a narrow Fe K emission with a center
energy of $\sim 6.40$ keV and $EW\sim 130$ eV (Yaqoob et al. 2001),
similar
to the average line's $EW$ we find for this object (Table 3). As these
authors comment, the line should be produced in material that is located
away from the central source, in agreement with our results from the
hardness ratios variations.

Our results do not exclude reprocessing from material close to the central
source (i.e. from the accretion disk). The $HR2$ color-flux diagram shows
significant scatter around the best fitting lines. These variations could
be the result of the slope variations that cause the scatter in the $HR1$
color-flux plot (the $[7-10]$ keV band count rate contributes to both the
$HR1$ and $HR2$ ratios). However, the $HR2$ variations at constant flux
could also be the result of fast variations of the iron line flux.

Finally, we find a correlation between the spectral parameters of the
average spectra of the individual sources in the sense that softer spectra
show a lager reflection component $R$ and line's $EW$ value. This is
similar to the correlation between $R$ and $\Gamma$ of Zdziarski et al.
(1999) for different sources. However, we do not find the same correlation
between $R$ and $\Gamma$ during the spectral variations within each
source. Our results show clearly that the reflection parameter $R$ is not
increasing with increasing source flux (i.e. as the spectrum softens). The
$R$-flux correlation that we find is opposite to what we would expect if
$R$ was decreasing with decreasing $\Gamma$. Therefore, we do not confirm
the Zdziarski et al. (1999) correlation between $\Gamma - R$ during the
spectral variations within each source. If the bulk of reprocessing
originates from neutral material located away from the central source {\it
and} the $\Gamma-R$ relationship is valid for different sources, then its
origin is not obvious. It probably depends on a physical parameter that
varies from one object to the other, like black hole mass or luminosity
(which depends on the accretion rate as well). For example, in the more
luminous sources the flux of the soft photons could be stronger resulting
in a stronger cooling of the plasma hence a softer spectrum. At the same
time, the effective solid angle subtended by the reflector should also be
larger, either because the reflector is located closer to the central
source (which seems unlikely), or because the reflector is having a larger
size.

\section{Conclusions}

We have used monitoring $RXTE$ light curves to study the spectral
variability in four Seyfert galaxies, namely NGC~4051, MCG~-6-30-15,
NGC~5506 and NGC~5548. The large number of observations over a long period
(over $\sim 3$ years) allowed us to study their overall spectral
variability behavior over many, different flux states. This is not
possible to achieve with the normal observations of AGN which last
typically less than a few days.

The exposure time of each observation is not long enough to perform model
fitting to the $[3-15]$ keV band energy spectrum of the sources, but
adequate to give us accurate measurements of the flux in different energy
bands. For that reason, we computed hardness ratios, we produced
color-flux diagrams and we used the commonly used ``power law plus
reflection plus iron line" model to interpret our results.

All sources show similar spectral variations. The variations in the $HR1$
color-flux plot imply primary slope changes which affect significantly the
soft band flux.  The mean trends in the $HR2$ and $HR3$ color plots can be
explained if we assume that the flux of the line and of the reflection
component remains roughly constant during the underlying continuum
variations. A constant line and reflection component flux is naturally
expected if the reprocessing material is located away from the central
source. The fact that the model under the assumption of constant line and
reflection flux describes well the mean trend in the color-flux of all
sources suggests that they operate in the same way. What are different is
the amplitude of the slope variations (NGC 4051 shows the largest
amplitude slope variability) and the average spectrum of each source.
Interestingly, we find that sources with softer energy spectra show larger
$EW$ and $R$ values as well. However, within each source, we find the
opposite behavior. The reflection parameter $R$ and the line's $EW$
decrease (instead of increasing) with increasing source flux (and hence
softening of the energy spectrum). Therefore, we confirm the $R-\Gamma$
relationship of Zdziarski et al. (1999) for the average spectra of the
sources, but not for the spectral variations within each source.

The significant scatter around the best fitting lines in the $HR1$ and
$HR2$ color-flux diagrams shows that the spectral variability in AGN is
more complicated than the mean trends of the color-flux plots suggest. The
low amplitude slope variations at constant flux imply that Compton-cooling
is not the only process that determines the properties of the active
corona. The $HR2$ variations around the mean trend could imply low
amplitude, line flux variations. If true, they could be the result of
reprocessing from material near the central source.

\acknowledgements
We thank the {\it RXTE} team for their operation of the satellite. IEP
thanks the Osservatorio Astronomico di Brera for hospitality. We also
thank an anonymous referee for making helpful suggestions. Part of this
work was done in the TMR research network 'Accretion onto black holes,
compact stars and protostars' funded by the European Commission under
contract number ERBFMRX-CT98-0195.

{\bf APPENDIX}

FALSE CORRELATIONS IN THE COLOR-COLOR PLOTS

As stated in the main text, the use of the same count rate in the
determination of two colors can introduce misleading correlations in the
color-color diagrams. To illustrate this, in the top plot in Figure 8 we
have plotted the $HR2$ versus $HR1$ colors in the case of MCG~-6-30-15.
There appears to exist a strong correlation between the two ratios in the
sense that $HR2$ increases with increasing $HR1$. The solid line in this
plot shows the best fitting line to the color-color diagram (we find a best
fitting slope of $1.24\pm 0.09$). This is not consistent with the $HR2$ vs
$C_{(3-5keV)_{norm}}$ color-flux diagram of the same source which showed
that $HR2$ remains constant, independent of the source flux state. Since
the spectral slope is correlated with the source flux, we did not expect to
see a correlation between $HR2$ and $HR1$ either.

We performed a numerical experiment to verify whether the observed
correlation between $HR2$ and $HR1$ is an artifact, caused by the
inter-dependence of the two variables. First, we assumed 124 points
(equal to the number of points in the MCG~-6-30-15 light curve) evenly
distributed in the interval between $0.4$ and $1.8$ (i.e. the limits of
the $C_{(3-5keV)_{norm}}$ values of MCG~-6-30-15). Using the best
fitting power law model line of the $HR1$ vs $C_{(3-5keV)_{norm}}$
color-flux diagram, we computed the $HR1$ values that correspond to
these values, and hence the respective $[7-10]$ keV band count rate. To
compute the $C_{5-7keV}$ keV values, we used the $C_{7-10keV}$ values
and assumed a constant $HR2$ value (equal to the mean $HR2$ value listed
in Table 1) at each flux state. Then we added appropriate random
Gaussian ``errors" to the simulated $C_{3-5keV}$, $C_{5-7keV}$ and
$C_{7-10keV}$ points and computed the $HR1$ and $HR2$ colors using the
randomized simulated count rates.

The simulated $HR1$ and $HR2$ vs. $C_{(3-5keV)_{norm}}$ color-flux
diagrams are shown in Figure 8 (second and third plot from top). They
are similar to the respective diagrams in Figures 3 and 4. The bottom
plot in Figure 8 shows the $HR2$ vs $HR1$ diagram using the simulated
colors. Despite the fact that we had computed the $C_{5-7keV}$ simulated
values in such a way so that $HR2$ would remain constant at the same
value irrespective of the source flux or spectral state, the simulated
color-color plot shows a strong correlation between $HR2$ and $HR1$, as
if larger values of $HR2$ correspond to larger values of $HR1$ as well.
In fact, the slope of the best fitting line to the simulated color-color
plot (shown with a solid line in Figure 8) is $1.49\pm 0.17$. The
difference between this value and the best fitting slope value of the
observed color-color diagram is $0.25\pm 0.17$. This result shows that
purely statistical effects are highly likely to reproduce the observed
correlation between $HR1$ and $HR2$. Since both ratios use $C_{7-10keV}$
in the numerator, when $C_{7-10keV}$ is higher than average due to a
statistical fluctuation, then both $HR1$ and $HR2$ will be higher on
average, resulting in an almost one-to-one correlation between them.

\clearpage

\figcaption{The {\it RXTE} $2-10$ keV monitoring light curves of
NGC~4051, MCG~-6-30-15, NGC~5506 and NGC~5548. The light curves are
normalized to their weighted mean. The time axis is on days since the
first observation of each source. Note the different sampling scheme
throughout the observing period. Errors are included in the plotted points
but are too small to be seen.}

\figcaption{ The hardness ratio light curves for the four sources. The
solid lines show the mean value of each light curve. Time is measured (in
days) since first observation of each source. Note the different $y-$axis
scaling in the case of NGC 4051.}

\figcaption{The $HR1$ values plotted as a function of the normalized
$[3-5]+[7-10]$ keV source count rate. The dashed lines show the best
fitting power law to the data. Filled squares show the slope of a power
law plus reflection model that corresponds to the respective [flux,
$HR1$] values. These slope values were computed as described in Section
4.1}

\figcaption{The $HR2$ values plotted as a function of the normalized
$[3-5]$ keV source count rate. Filled squares show the average $HR2$
values in various flux bins. In all plots, the dashed, dotted and
dashed-dotted lines show the color-flux model curves in the case of a
power law plus reflection plus iron line with constant $EW=0, 250$ and
$500$ eV respectively.  Solid lines show the model color-flux curves in
the case of a constant flux line. The model curves were computed as
explained in Section 4.1}

\figcaption{The $HR3$ values plotted as a function of the normalized
$[3-5]$ keV source count rate.  Filled squares show the average $HR3$
values in various flux bins. In all plots, the dashed, dotted and
dashed-dotted lines show the color-flux model curves in the case a power
law plus reflection with constant $R=2, 1$ and $0$ respectively. Solid
lines show the model color-flux curves in the case of a constant flux
reflection. The model curves were computed as explained in Section 4.1}

\figcaption{The $HR1$ values plotted as a function of the normalized
$[3-5]+[7-10]$ keV source count rate. As in Figure 3, the dashed lines
show the best fitting power law to the data. Filled squares show the slope
of the power law that corresponds to the respective [flux, $HR1$] values
in the case when there exists a reflection component whose flux remains
constant. These values were computed as described in Section 4.2.2. Note
that, in this Figure, we have included the data from the low state of
NGC~4051.}

\figcaption{Plot of the model, $[1-300]$ keV flux as a function of the
normalized, soft band count rate. The peak-to-peak variation in the source
luminosity is less than $\sim 2$ in all cases.} 

\figcaption{Plot of the $HR2$ vs $HR1$ color-color diagram in the case of
MCG~-6-30-15 (a). A strong correlation is observed. The solid line shows
the best fitting line to the data. The panels (b) and (c) show a plot of
the $HR1$ and $HR2$ vs normalized [$3-5$] keV count rate color-flux
diagrams for a set of simulated light curves (see Appendix). Panel (d)
shows the $HR2$ vs $HR1$ color-color diagram for the same set of simulated
light curves. Although the light curves were constructed in such a way so
that $HR2$ is constant, there exists a strong correlation between $HR2$
and $HR1$ (similar to the correlation between the observed $HR1$ and $HR2$
values of MCG~-6-30-15 shown in panel (a)). This correlation is caused by
random, statistical effects.}

\newpage
\plotone{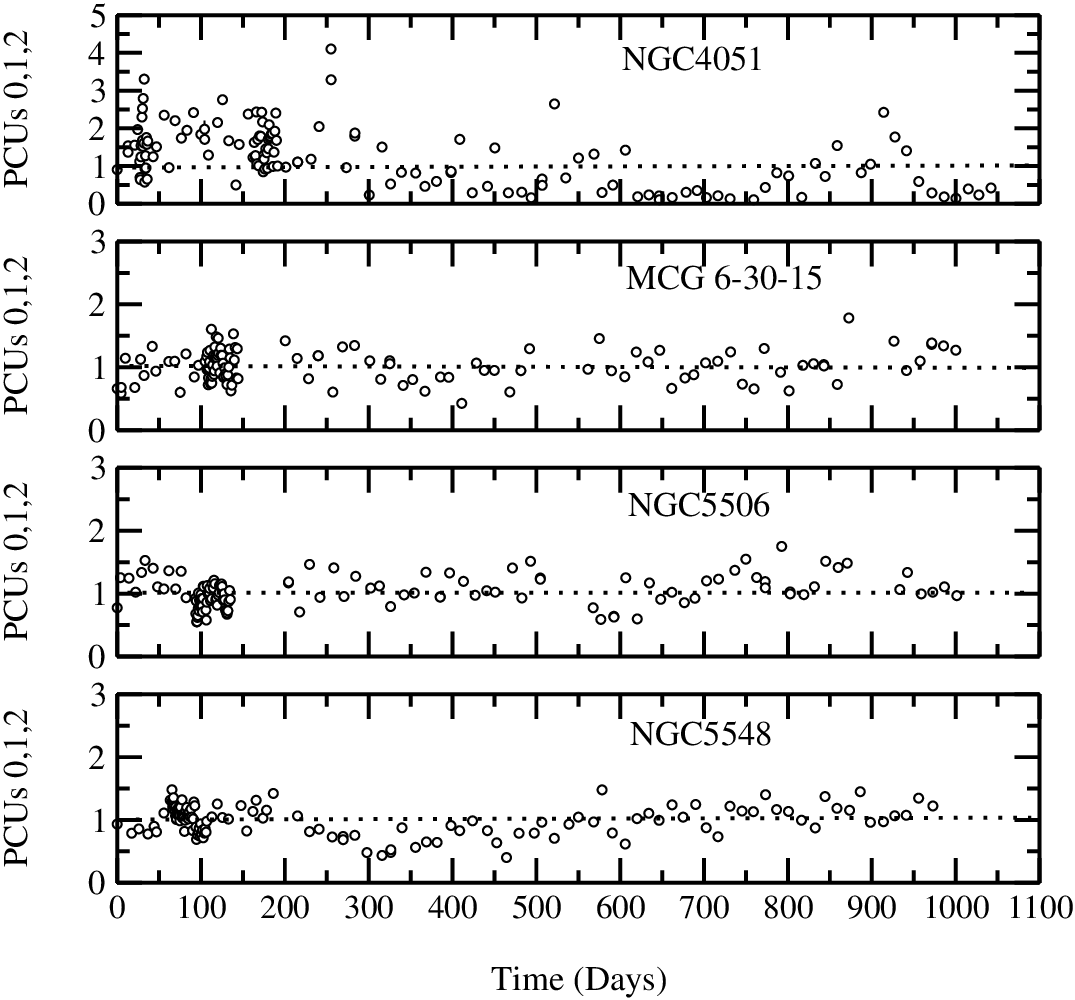}
\newpage
\plotone{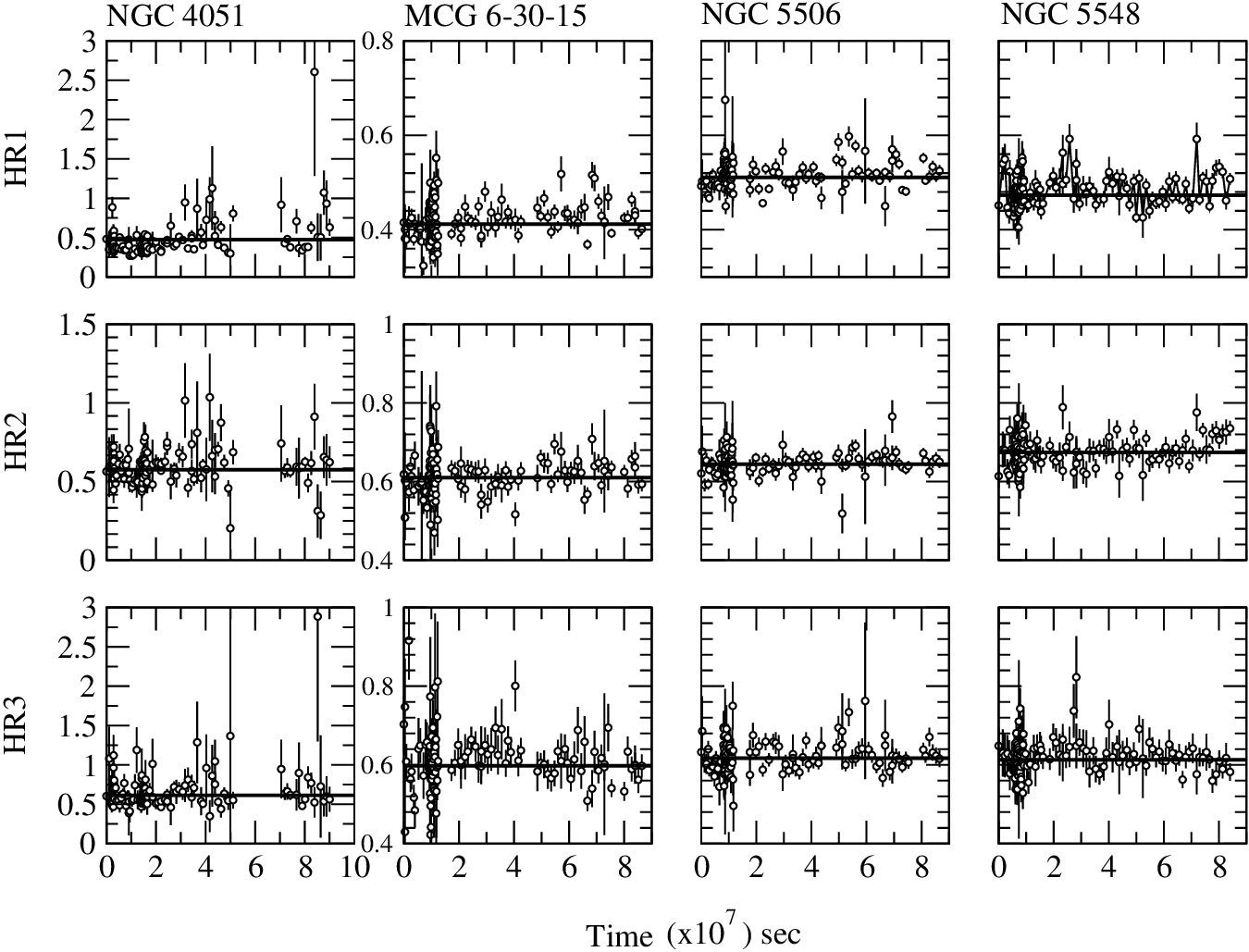}
\newpage
\plotone{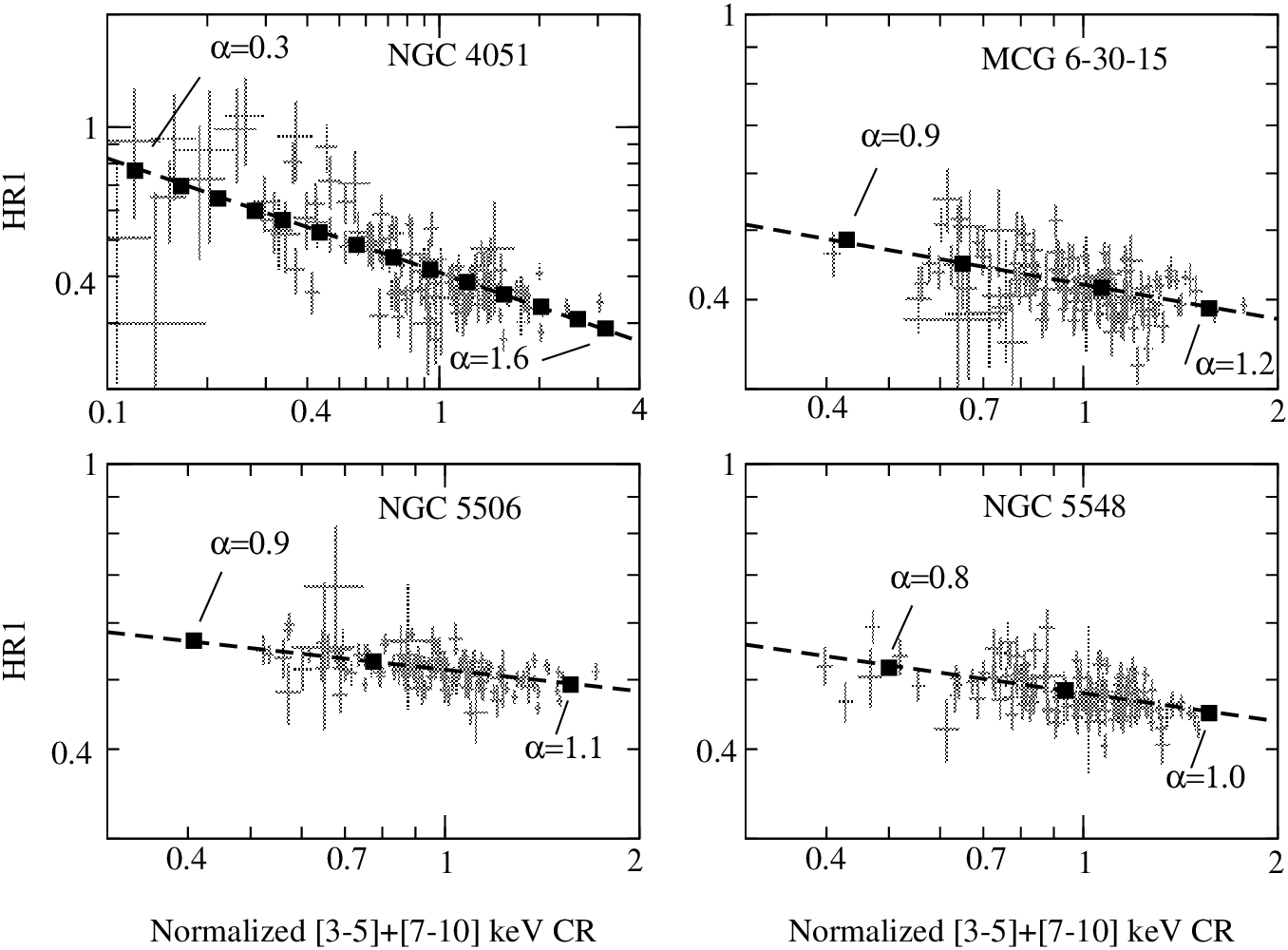}
\newpage
\plotone{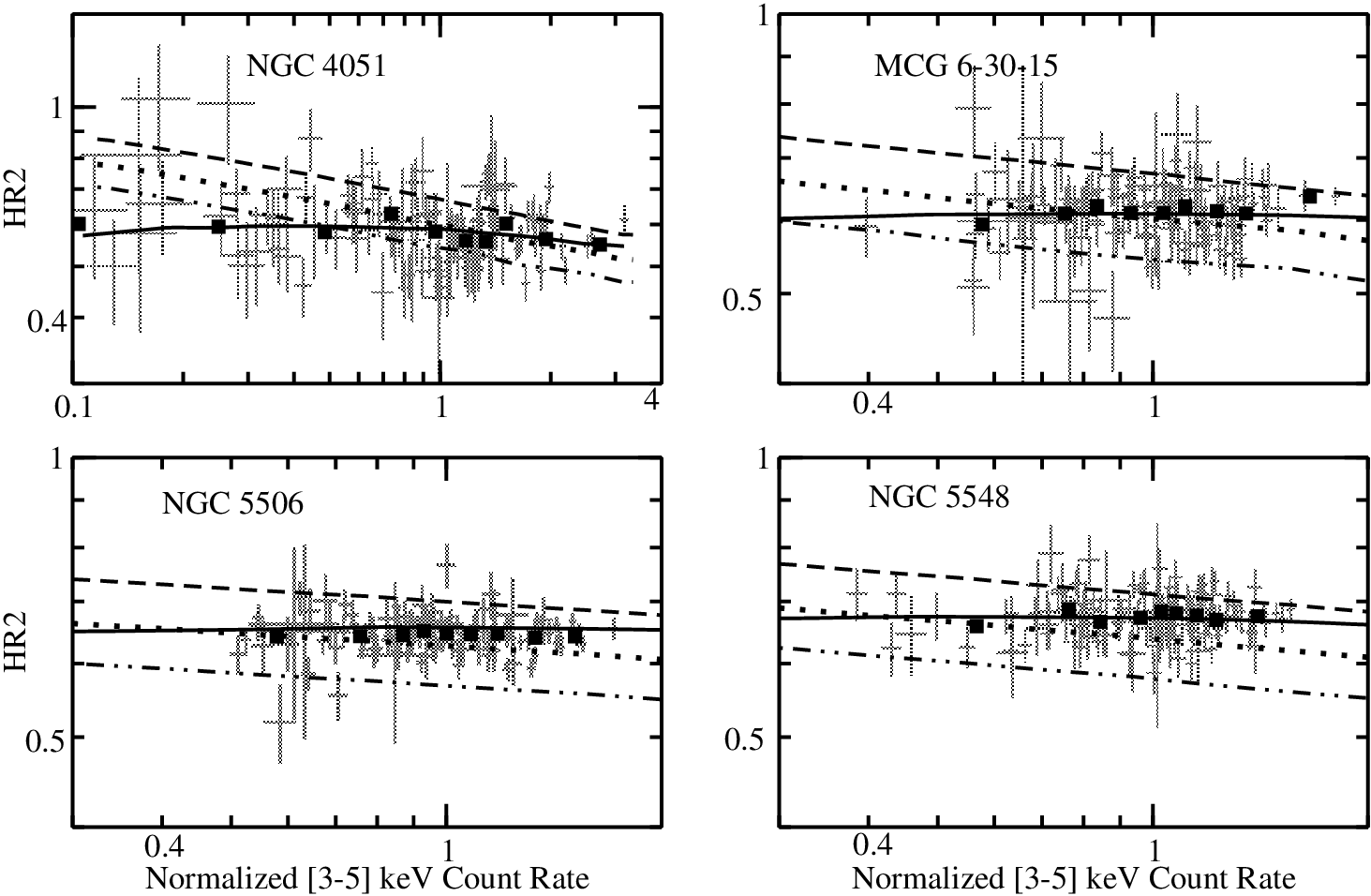}
\newpage
\plotone{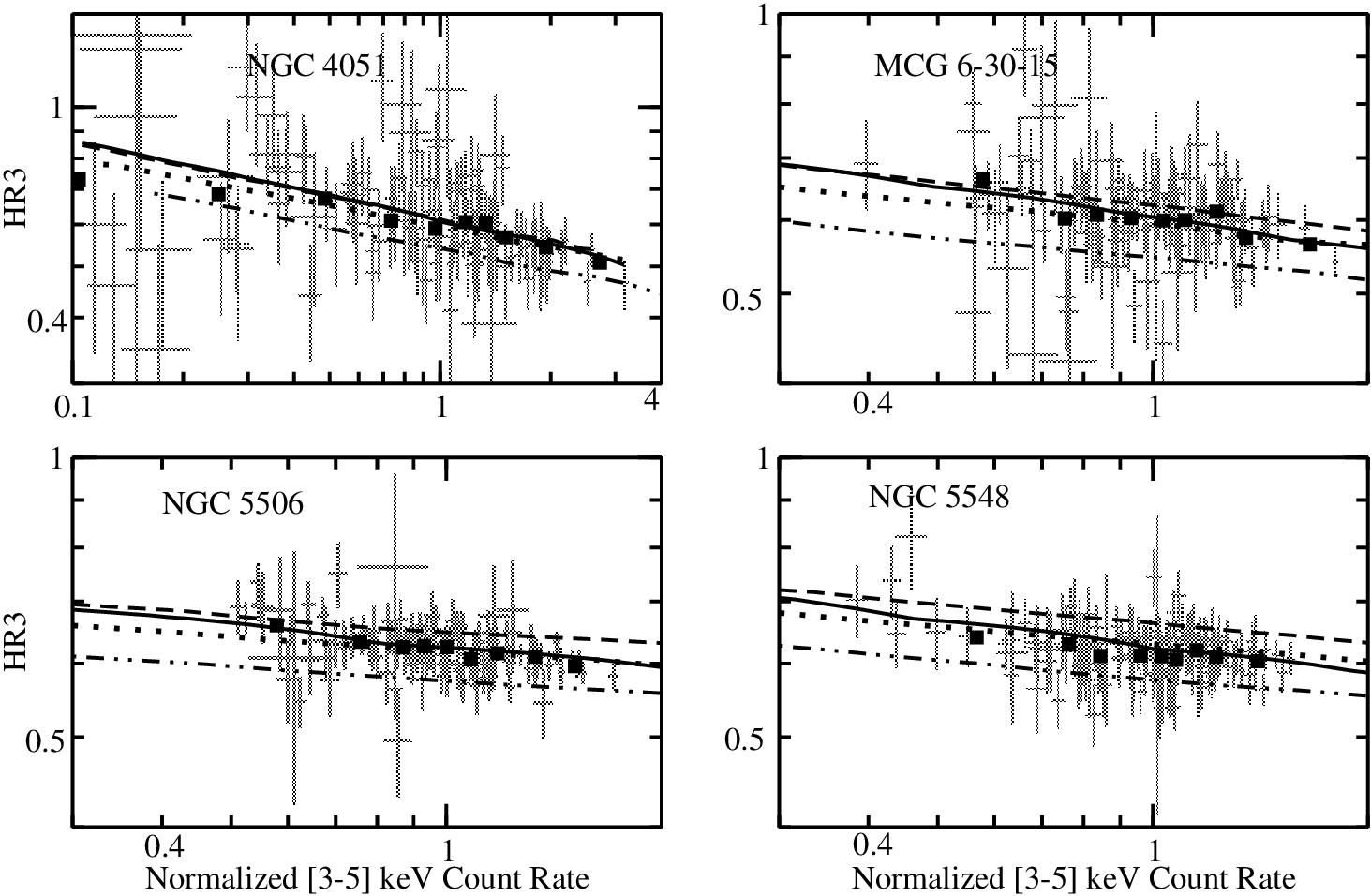}
\newpage
\plotone{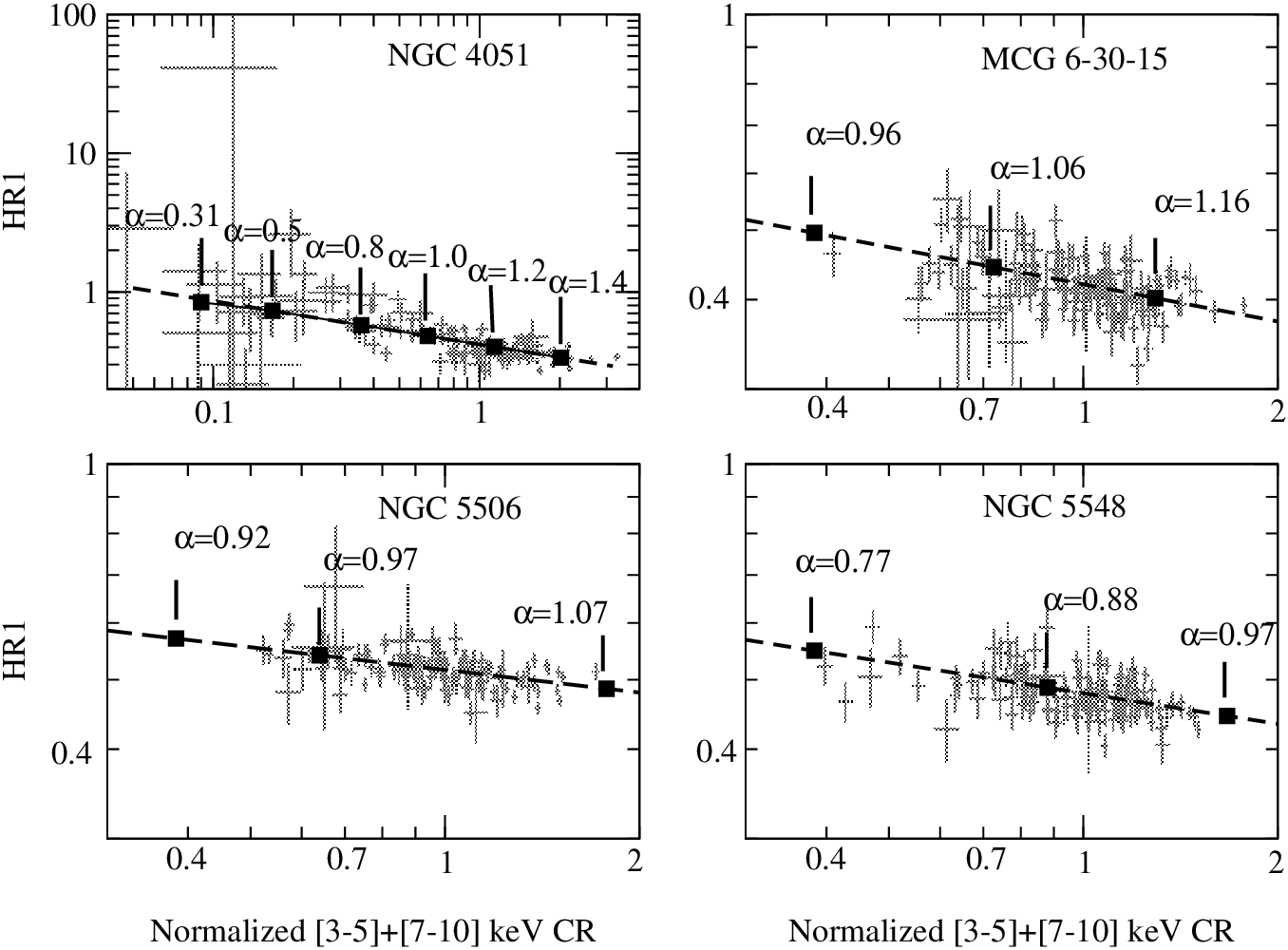}
\newpage
\plotone{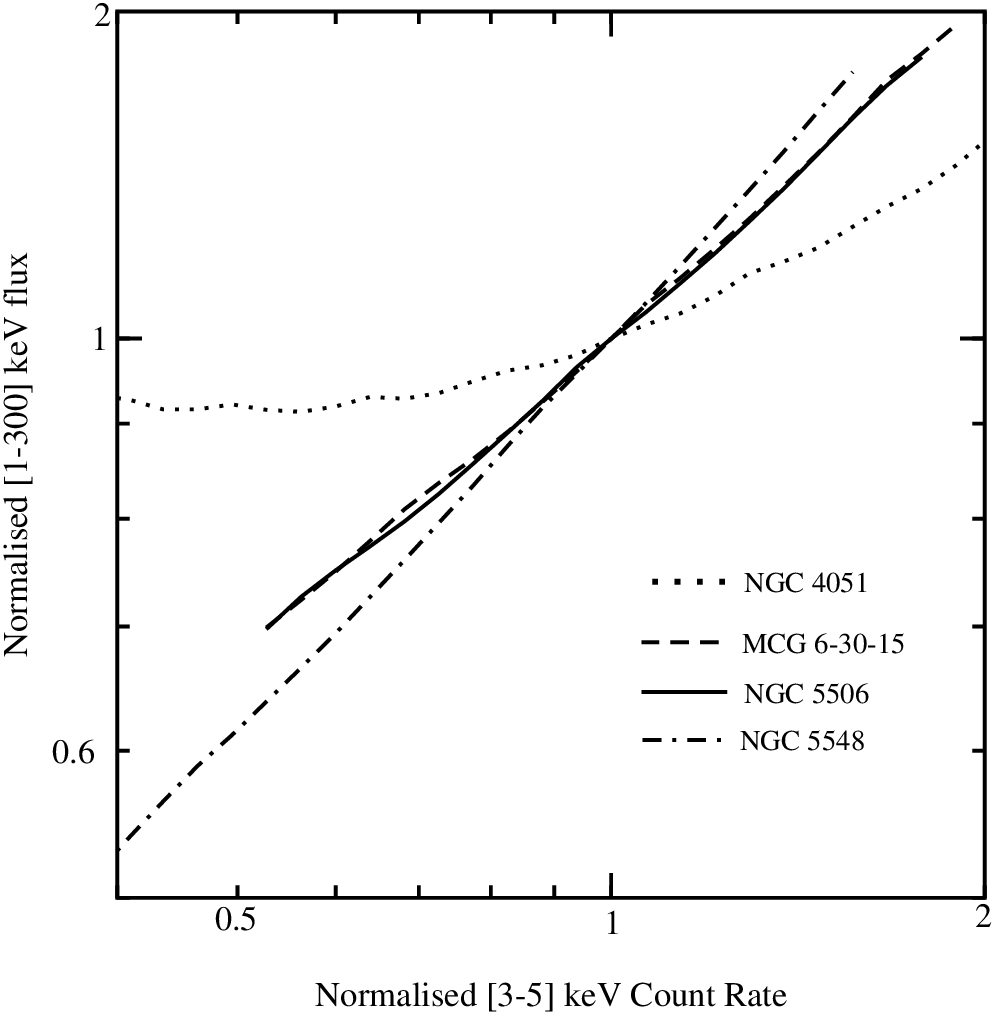}
\newpage
\plotone{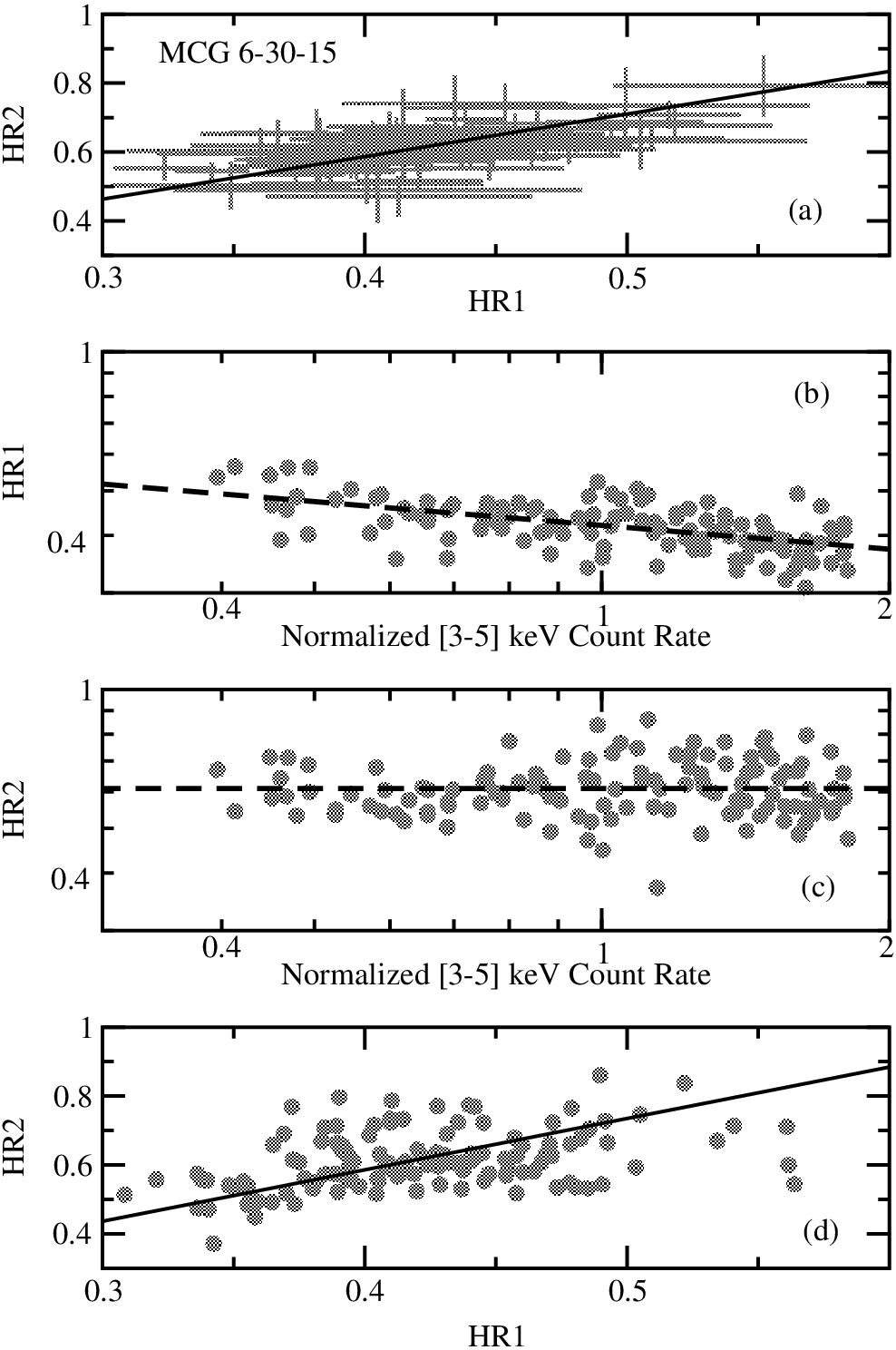}

\clearpage

\begin{table}[t]
\begin{tabular}{ccccccc}
\hline
\hline
Name & $\bar{HR1}$ & $\sigma_{rms}$ &$\bar{HR2}$
& $\sigma_{rms}$ &$\bar{HR3}$ & $\sigma_{rms}$ \\
\hline
NGC~4051 & 0.37 & 0.17 & 0.57 & 0.06 & 0.58 & 0.02 \\
MCG~-6-30-15 & 0.41 & 0.06 & 0.61 & 0.04 & 0.60 & 0.05 \\
NGC~5506 & 0.51 & 0.03 & 0.64 & 0.03 & 0.62 & 0.04 \\
NGC~5548 & 0.47 & 0.04 & 0.67 & 0.02 & 0.61 & -   \\
\hline
\end{tabular}
\caption{The mean and $\sigma_{RMS}$ values of the $HR1$, $HR2$, and
$HR3$ light curves shown in Figure~2. The error on the mean $HR$ values is
less than $0.01$. }
\label{tab1}
\end{table}


\begin{table}[t]
\begin{tabular}{lccc}
\hline
\hline
Object Name & $HR1/C_{(3-5keV+7-10keV)_{norm}}$ &
$HR2/C_{(3-5keV)_{norm}}$
& $HR3/C_{(3-5keV)_{norm}}$  \\
  & $b_{1}$ $(\chi^{2}/$dof) & $b_{1}$ $(\chi^{2}/$dof) &
$b_{1}$ $(\chi^{2}/$dof)  \\
\hline
NGC~4051 & $-0.30\pm0.02$(284/123) & $-0.05\pm0.01$(189/123) & $-0.17\pm
0.02$(124/123) \\
MCG~-6-30-15 & $-0.16\pm0.02$(256/122) & $+0.02\pm0.02$(190/122) &
$-0.13\pm0.03$(132/122) \\
NGC~5506 & $-0.10\pm0.01$(293/125) & $-0.01\pm0.01$(193/125) & $-0.10\pm
0.01$(130/125) \\
NGC~5548 & $-0.13\pm0.01$(290/132) & $+0.01\pm0.01$(191/132) & $-0.06\pm
0.02$(110/132) \\
\hline
\end{tabular}   
\caption{Best power law model fitting results to the color-flux plots
shown in Figures 3, 4 and 5.}
\label{tab2}
\end{table}

\clearpage

\begin{table}[t]
\begin{tabular}{cccc}
\hline
\hline
Name & $\langle\Gamma\rangle$ & $\langle R \rangle$ & $\langle EW\rangle$
(eV) \\
\hline
NGC~4051 & $2.06$ & $1.7$ & $310$\\
MCG~-6-30-15 & $2.05$ & $1.2$ & $180$ \\
NGC~5506 & $2.11$ & $1.4$ & $230$ \\
NGC~5548 & $1.88$ & $0.8$ & $130$ \\
\hline
\end{tabular}
\caption{ The mean spectral index, reflection normalization $R$ and iron
line equivalent width ($EW$) of the sources in our sample, computed
using the mean hardness ratio values and a power law plus reflection and
a narrow Gaussian line spectral model (see Section 4.2.1).}
\end{table}


\begin{table}[t]
\begin{tabular}{ccccc}
\hline
\hline
Name & $HR2/C_{(3-5keV)_{norm}}$ & & $HR3/C_{(3-5keV)_{norm}}$ &  \\
  & $b_{CEW}$ & $b_{CLF}$ & $b_{CR}$ & $b_{CRF}$ \\
\hline
NGC~4051 & -0.14 & -0.03 & -0.13 & -0.15 \\
MCG~-6-30-15 & -0.08 & -0.02 & -0.08 & -0.11 \\
NGC~5506 & -0.04 & -0.02 & -0.05 & -0.09 \\ 
NGC~5548 & -0.06 & -0.03 & -0.06 & -0.09 \\
\hline
\end{tabular}
\caption{ The slope of the model curves shown in Figures 4 and 5.
$b_{CEW}$ and $b_{CLF}$ represent the slope of the model curves under the
assumption of constant line's $EW$ and flux respectively. Similarly,
$b_{CR}$ and $b_{CRF}$ denote the slope of the model curves in the case of
constant $R$ and constant reflection flux respectively. }
\end{table}

\end{document}